\documentclass{aastex631}

\usepackage{lastpage} 
\usepackage{booktabs}  
\usepackage{graphicx}  
\usepackage{longtable} 
\usepackage{lipsum} 
\usepackage{array} 
\usepackage{hyperref}   
\usepackage{microtype}
\usepackage{graphicx} 
\usepackage{xcolor} 

\usepackage[T1]{fontenc}
\usepackage{booktabs, xcolor, url}
\definecolor{sncolor}{RGB}{255,245,230}
\definecolor{kncolor}{RGB}{230,240,255}

\begin{document}

\title{Spectral Hardness as the Primary Discriminator: Unveiling the Collapsar--Merger Boundary with a Gold-Standard GRB Sample
}

\correspondingauthor{Yan-Kun Qu}
\email{quyk@qfnu.edu.cn}

\author{Xue Zhang}
\affiliation{School of Physics and Physical Engineering, Qufu Normal University, Qufu 273165, People's Republic of China}

\author[0000-0002-9838-4166]{Yan-Kun Qu}
\affiliation{School of Physics and Physical Engineering, Qufu Normal University, Qufu 273165, People's Republic of China}
\affiliation{Guangxi Key Laboratory for Relativistic Astrophysics, School of Physical Science and Technology, Guangxi University, Nanning 530004, People’s Republic of
China}

\author[0000-0003-0672-5646]{Shuang-Xi Yi}
\affiliation{School of Physics and Physical Engineering, Qufu Normal University, Qufu 273165, People's Republic of China}

\author[0000-0002-0786-7307]{Yu-Peng Yang}
\affiliation{School of Physics and Physical Engineering, Qufu Normal University, Qufu 273165, People's Republic of China}

\author[0000-0002-6072-3329]{Fen Lyu}
\affiliation{School of Physics and Astronomy, Anqing Normal University, Anqing 246133, People's Republic of China}
\affiliation{Institute of Astronomy and Astrophysics, Anqing Normal University, Anqing 246133, People's Republic of China}

\author{Fa-Yin Wang} 
\affiliation{School of Astronomy and Space Science, Nanjing University, Nanjing 210093, People's Republic of China}
\affiliation{Key Laboratory of Modern Astronomy and Astrophysics, Nanjing University, Nanjing 210093, People's Republic of China}

\author{Zhong-Xiao Man}
\affiliation{School of Physics and Physical Engineering, Qufu Normal University, Qufu 273165, People's Republic of China}

\begin{abstract}
Gamma-ray bursts (GRBs) are promising cosmological probes. However, 
their utility is hampered by sample impurity arising from traditional 
duration-based classification. The existence of ``hybrid'' events---such as the merger-origin GRBs 060614 and 211211A that exhibit extended durations ($T_{90} \gg 2$\,s)---highlights the limitations of the $T_{90}$ parameter as a sole discriminator.
In this Letter, we establish a robust, physically motivated classification method using a Support Vector Machine (SVM) trained on a ``gold-standard'' sample of 24 GRBs with spectroscopically confirmed progenitors (associated SNe or KNe). By isolating the prompt main spike to excise contamination from extended emission, we derive a quantitative classification index, $I_{\rm SVM} = 5.01 \log_{10} E_{p,i} - 1.25 \log_{10} E_{\rm iso} - 0.34 \log_{10} T_{90,z} - 12.90$ (units: keV, $10^{52}$\,erg, s). Events with $I_{\rm SVM} > 0$ are classified as mergers. 
Analysis of the standardized classification weights reveals that the discriminative power of $E_{p,i}$ is approximately \textbf{5 times} that of $T_{90,z}$, while $E_{\rm iso}$ contributes a weight comparable to $E_{p,i}$. This quantitatively demonstrates that \textbf{spectral hardness and energetics}, rather than duration, are the primary physical signatures distinguishing mergers from collapsars.
The derived boundary implies a stringent hardness ceiling for collapsars, while mergers are identified as outliers with excessive hardness relative to their energy budget. 
The classifier successfully identifies the nature of historic test cases, including the ultra-long GRB 111209A (collapsar) and the short GRB 050709 (merger), independent of instrumental eras. This tool paves the way for cleaning archival and future high-redshift GRB samples for precision cosmology.
\end{abstract}


\section{Introduction}
\label{sec:intro}

As one of the most energetic transient phenomena in the universe, gamma-ray bursts (GRBs) have attracted significant attention in high-energy astrophysics. Their immense luminosity makes them detectable to very high redshift (e.g., $z > 8$) \citep{2009Natur.461.1258S, 2009Natur.461.1254T}. A long-standing goal in astrophysics is to use GRBs as ``standardizable candles'' to constrain cosmological parameters and to complement Type Ia supernovae (SNe Ia) \citep{Wang2015, 2017AdAst2017E...5C}. However, applications to large samples remain limited by substantial scatter in GRB luminosity correlations \citep{2013IJMPD..2230028A, 2022ApJ...924...97W, 2023ApJ...953...58L, 2024A&A...689A.165L, 2023ApJ...958...74T, 2025ApJ...991..145Z}. Recent studies identify \textit{sample impurity} as a significant and correctable source of systematic bias: the contamination of the long-GRB population by merger-origin events introduces systematic biases into luminosity correlations, hampering their utility for precision cosmology \citep{2006Natur.444.1050D, 2024ApJ...968...38D, 2024ApJ...976..170Q, 2025MNRAS.540L...6Q}.

Traditionally, GRBs are classified into long ($T_{90} > 2$\,s) and short ($T_{90} < 2$\,s) populations based on the bimodal duration distribution \citep{1993ApJ...413L.101K, 2013ApJ...764..179B}. This dichotomy was originally thought to map directly onto distinct physical progenitors: Type II GRBs, originating from the core collapse of massive stars (Collapsars), and Type I GRBs, resulting from compact object mergers (Mergers) \citep{2006ApJ...642..354Z, 2009ApJ...703.1696Z}. Collapsars are typically associated with Type Ic supernovae and are found in star-forming galaxies \citep{1998Natur.395..670G, 2006ARA&A..44..507W}. In contrast, mergers are linked to kilonovae (KNe) and gravitational waves (e.g., GRB 170817A; \citealt{2017PhRvL.119p1101A}), often occurring in environments with older stellar populations \citep{2007PhR...442..166N}.

However, the correlation between duration and progenitor type is far from straightforward. The observed $T_{90}$ is heavily influenced by detector sensitivity, energy band, and viewing geometry, differing from the intrinsic engine timescale by up to an order of magnitude \citep{2013ApJ...765..116K, 2022ApJ...927..157M, 2025JHEAp..45..325Z}. The limitations of duration-based classification have become evident with the discovery of ``hybrid'' events. Peculiar long-duration bursts such as GRB 060614, GRB 211211A, and GRB 230307A have been confirmed to originate from mergers (associated with KNe), yet their prompt emission was extended by softer components or magnetic braking (often manifesting as extended emission), mimicking collapsars \citep{2006Natur.444.1050D, 2022Natur.612..223R, 2024Natur.626..737L}. Conversely, the short-duration GRB 200826A ($T_{90} \sim 1$\,s) was confirmed to be a collapsar driving a supernova \citep{2021NatAs...5..911Z}. These outliers suggest that a significant fraction of low-redshift long GRBs may originate from misclassified merger events rather than massive star collapse.

Identifying the physical origin of these bursts is critical but challenging. While the detection of a supernova or kilonova provides a smoking gun, such signatures are intrinsically faint. Unambiguous spectroscopic confirmation of the associated supernova becomes increasingly difficult beyond $z \sim 1$, even with the sensitivity of the \textit{James Webb Space Telescope} (JWST), as the SN component is often outshone by the bright afterglow or overwhelmed by the host galaxy \citep{2006SSRv..123..485G}. In contrast, advanced missions like \textit{SVOM} \citep{2016arXiv161006892W} and the \textit{Einstein Probe} \citep{2022hxga.book...86Y} are expected to detect GRBs out to $z \sim 15$ \citep{2024A&A...685A.163L}. For these high-redshift events, classification must rely solely on prompt emission parameters. Previous multi-parameter schemes, such as the $\epsilon$ parameter ($\epsilon = E_{\gamma,\rm iso} / E_{p}^{5/3}$) proposed by \cite{2010ApJ...725.1965L}, marked a significant step forward. However, these methods were often calibrated on statistical samples that inevitably contained misclassified events, limiting their precision in the era of multi-messenger astronomy.

In this work, we establish a robust, physically motivated classification method by prioritizing sample purity over quantity. We train a Support Vector Machine (SVM) on a ``gold-standard'' dataset of 24 GRBs with spectroscopically confirmed progenitors (17 collapsars, 7 mergers). By explicitly isolating the prompt main spike to excise contamination from extended emission, the resulting decision boundary in the multi-dimensional $E_{p,i}-E_{\rm iso}-T_{90,z}$ parameter space effectively identifies the intrinsic physical signatures of these events. This provides a practical tool for distinguishing progenitor types in the vast population of archival and future GRBs.

The paper is organized as follows: Section \ref{sec:sample} describes the selection of the gold-standard sample; Section \ref{sec:method} details the feature selection and SVM algorithm; Section \ref{sec:Results} presents the classification boundary and validation results; and Section \ref{sec:discussion} discusses the physical implications and future applications.

\section{Sample Selection}
\label{sec:sample}

Unlike statistical studies that maximize sample size, our machine learning approach prioritizes sample purity. High-fidelity training labels are essential to delineate the intrinsic physical boundary between progenitor types. Consequently, we construct a ``gold-standard'' dataset comprising \textbf{17} collapsar-origin GRBs (sn-GRBs) and \textbf{7} merger-origin GRBs (kn-GRBs), selected based on spectroscopic confirmation of associated transients (SNe or KNe).

\subsection{The Collapsar Sample (sn-GRBs)}

Although the connection between long GRBs and core-collapse supernovae (SNe) was established over two decades ago \citep{1998Natur.395..670G}, confirmed associations remain rare among the $\sim$2000 bursts detected in the \textit{Swift} era. We adopt the classification scheme of \cite{2012grb..book..169H}, restricting our sample to events with strong spectroscopic evidence (Class A) or a clear light curve bump compatible with a GRB-SN spectrum (Class B). 

To ensure a physically and instrumentally homogeneous training set, we apply the following rigorous selection criteria:

\begin{enumerate}
    \item \textbf{Instrumental Homogeneity:} 
    To minimize systematic uncertainties arising from cross-calibration between detector generations, we strictly limit our sample to events detected after the launch of \textit{Swift} (2004). This ensures that spectral parameters (particularly $E_{p}$) are derived from modern high-sensitivity instruments. Consequently, historic pre-Swift events such as GRB 030329 are excluded 
from training and reserved as independent test cases to validate 
cross-era generalization.

    \item \textbf{Physical Purity (Jet-powered events only):} 
    We exclude Low-Luminosity GRBs (llGRBs) such as GRB 060218 (and the pre-\textit{Swift} GRB 980425). These events exhibit distinct single-peaked light curves and soft spectra likely originating from shock breakouts rather than fully developed relativistic jets \citep{2001ApJ...551..946T, 2015ApJ...807..172N}, representing a distinct physical channel that would contaminate the classifier.
    
    \item \textbf{Typical Collapsar Regime:} 
    We exclude ultra-long bursts, specifically GRB 111209A. Its extreme duration and exceptional SN luminosity ($>5\sigma$ above average) suggest a central engine regime (e.g., magnetar-driven) distinct from typical collapsars \citep{2015Natur.523..189G}. Like historic bursts, GRB 111209A is reserved to test the classifier's generalization to extreme durations.
    
    \item \textbf{Duration-independent Training:} 
    To prevent the SVM from simply reproducing the traditional $T_{90}$ cut, we explicitly include the ``short'' collapsar \textbf{GRB 200826A}. Despite its duration of $T_{90} < 2$\,s, it is spectroscopically confirmed as a collapsar \citep{2021NatAs...5..911Z}. Including this boundary case is critical for training the model to recognize collapsars based on spectral physics rather than duration alone.
\end{enumerate}

Applying these criteria yields a final, high-fidelity sn-GRB training sample of \textbf{17 events} (see Table \ref{table1}).

\subsection{The Merger Sample (kn-GRBs)}

The sample of confirmed compact object mergers is intrinsically limited. We strictly select events based on multi-messenger or multi-wavelength evidence: (i) association with gravitational waves, (ii) spectroscopic identification of r-process elements, or (iii) high-confidence kilonova signatures (IR excess and color evolution) with supernovae ruled out.

Our selection includes the landmark multi-messenger event GRB 170817A \citep{2017ApJ...848L..12A,2017PhRvL.119p1101A} and the recent bright burst GRB 230307A, which exhibited telltale spectral features of Tellurium \citep{2024Natur.626..737L}. We also include events with strong kilonova candidates such as GRB 130603B \citep{2013Natur.500..547T}, GRB 160821B \citep{2019MNRAS.489.2104T}, and the peculiar long-duration merger GRB 211211A \citep{2022Natur.612..223R}.

Consistent with the instrumental criteria applied to the collapsar sample, the classical short burst GRB 050709 (detected by HETE-2) is excluded from the training set \citep{2005Natur.437..845F,2005Natur.437..859H}. We reserve GRB 050709 as an independent test case to validate the model's generalizability to non-\textit{Swift}/\textit{Fermi} data. The final training set comprises 7 kn-GRBs~(see Table \ref{table1}).

\subsection{Parameter Extraction: Isolating the Main Spike}
\label{sec:main_spike}

A critical criterion for our ``gold-standard'' dataset is the physical isolation of the primary central engine activity. Hybrid merger events, such as GRB 060614 and GRB 211211A, frequently exhibit soft Extended Emission (EE) that can dominate the global duration and energy budget. Including the EE component would obscure the intrinsic properties of the relativistic jet launching phase, rendering the comparison with classical short GRBs physically invalid. 

Consequently, for these hybrid events, we did not use the global burst parameters; instead, we \textbf{explicitly adopted the spectral and temporal parameters of the isolated ``main spike'' component} as documented in detailed spectral analyses (\citealt{2006Natur.444.1044G, 2024ApJ...970....6X}; see Table \ref{table1} for complete references). For instance, for the canonical hybrid event GRB 060614, we adopted the main-spike duration of $T_{90,z} \simeq 4.4$\,s rather than its global duration of $>100$\,s. This selection strategy ensures that the input features for the SVM ($E_{p,i}$, $E_{\rm iso}$, $T_{90,z}$) consistently characterize the initial hard spike emission across both collapsar and merger populations.

To quantitatively assess the impact of this main-spike isolation criterion, we performed an ablation study by re-training the SVM using the time-integrated, full-emission parameters for the two hybrid events in our training set (GRBs 060614 and 211211A). Although only two events are altered, their hybrid nature makes them influential leverage points in the parameter space. The results indicate that incorporating the full EE systematically biases the learned hyperplane. For instance, using the isolated 4.44\,s main spike of GRB 060614 yields a robust merger classification ($P_{\rm sn} = 0.094$), whereas adopting its global 124\,s emission completely misclassifies it as a canonical collapsar. Furthermore, incorporating the full EE alters the global SVM weights: the coefficient for $\log_{10} E_{p,i}$ decreases from $5.01$ to $3.10$, the coefficient for $\log_{10} E_{\rm iso}$ shifts from $-1.25$ to $-1.11$, and notably, the weight for the duration term ($\log_{10} T_{90,z}$) flips from $-0.34$ to $+0.50$. This sign reversal suggests that the SVM is forced to use longer duration as a proxy for mergers, a behavior that is less physically meaningful and merely reflects the soft EE inflating the temporal baseline. Since EE likely arises from late-time central engine activity (e.g., fallback accretion or magnetar spin-down) rather than the initial prompt accretion phase, its inclusion inherently blurs the physical boundary between the two populations. Consequently, without the main-spike cut, the leave-one-out cross-validation (LOOCV) accuracy decreases to 83.3\%, leading to the misclassification of several canonical long SN-GRBs. This test supports our parameter extraction strategy, confirming that isolating the main spike is essential for uncovering the intrinsic physical boundary.

\begin{deluxetable*}{lcccccc}
\tablecaption{The Gold Standard Sample Parameters \label{table1}}
\tablewidth{0pt}
\tabletypesize{\small} 
\tablehead{
    \colhead{GRB} & 
    \colhead{$z$} & 
    \colhead{$T_{90,z}$} & 
    \colhead{Fluence} & 
    \colhead{$E_{p}$} & 
    \colhead{$E_{\gamma,\rm iso}$} & 
    \colhead{Refs.} \\
    \colhead{} & 
    \colhead{} & 
    \colhead{(s)} & 
    \colhead{($10^{-7}$ erg cm$^{-2}$)} & 
    \colhead{(keV)} & 
    \colhead{($10^{52}$ erg)} & 
    \colhead{}
}
\startdata
\cutinhead{KN Gold Sample (Mergers)}
060614  & 0.125   & 4.44   & $82^{+6}_{-25}$       & $302^{+214}_{-85}$ & $0.042^{+0.003}_{-0.013}$ & 1--3 \\
130603B                  & 0.360   & 0.13   & $66 \pm 7$            & $660 \pm 100$      & $0.22 \pm 0.02$           & 3--5 \\
150101B                  & 0.134   & 0.017  & $0.79 \pm 0.19$       & $125 \pm 49$       & $0.0004 \pm 0.0001$       & 3, 6, 7 \\
160821B                  & 0.160   & 0.41   & $1.67 \pm 0.25$       & $92 \pm 28$        & $0.0013 \pm 0.0002$       & 3, 6, 8 \\
170817A                  & 0.010   & 1.98   & $1.4 \pm 0.3$         & $215 \pm 54$       & $3.2 \times 10^{-6}$      & 3, 6, 7, 9 \\
211211A & 0.076   & 12.1   & $3770 \pm 10$         & $1360 \pm 78$      & $0.960 \pm 0.003$         & 10, 11 \\
230307A & 0.065   & 16.3   & $20300 \pm 30$        & $903 \pm 3$        & $3.260 \pm 0.005$         & 12, 13 \\
\cutinhead{SN Gold Sample (Collapsars)}
050525A                  & 0.606   & 5.50   & $153 \pm 2$           & $84 \pm 2$         & $2.20 \pm 0.03$           & 14--16 \\
081007                   & 0.529   & 6.54   & $12 \pm 1$            & $40 \pm 10$        & $0.25 \pm 0.02$           & 15, 17 \\
091127                   & 0.490   & 4.77   & $162 \pm 2$           & $59 \pm 2$         & $1.97 \pm 0.02$           & 6, 7, 15 \\
101219B                  & 0.550   & 21.9   & $22 \pm 1$            & $83 \pm 5$         & $0.172 \pm 0.009$         & 6, 7, 15 \\
120422A                  & 0.283   & 4.17   & $4.6^{+1.3}_{-1.7}$   & $24 \pm 10$        & $0.096^{+0.028}_{-0.036}$ & 15, 18 \\
130215A                  & 0.597   & 41.1   & $202 \pm 5$           & $102^{+102}_{-34}$ & $2.17 \pm 0.05$           & 15, 18 \\
130427A                  & 0.340   & 122    & $14200 \pm 20$        & $867 \pm 5$        & $60.00 \pm 0.07$          & 6, 15 \\
130702A                  & 0.145   & 51.5   & $63 \pm 20$           & $15.2 \pm 0.3$     & $0.032 \pm 0.010$         & 6, 7, 15 \\
130831A                  & 0.480   & 22.0   & $89 \pm 4$            & $54^{+7}_{-9}$     & $0.97 \pm 0.04$           & 14, 15 \\
141004A                  & 0.537   & 2.55   & $7.9^{+1.0}_{-0.8}$   & $64^{+12}_{-7}$    & $0.098^{+0.014}_{-0.009}$ & 15,19 \\
150818A                  & 0.282   & 36.7   & $53 \pm 7$            & $100^{+29}_{-18}$  & $0.16 \pm 0.02$           & 15, 20 \\
161219B                  & 0.147   & 6.05   & $27^{+3}_{-2}$        & $94^{+19}_{-12}$   & $0.022 \pm 0.002$         & 15, 19 \\
171010A                  & 0.329   & 80.7   & $6430 \pm 20$         & $194 \pm 1$        & $19.10 \pm 0.05$          & 6, 15 \\
180728A                  & 0.117   & 7.77   & $405 \pm 15$          & $88 \pm 9$         & $0.33 \pm 0.01$           & 15, 18 \\
181201A                  & 0.450   & 119    & $1990 \pm 60$         & $152 \pm 6$        & $11.9 \pm 0.4$            & 21, 22 \\
190829A                  & 0.079   & 54.0   & $116 \pm 3$           & $8.5 \pm 0.3$      & $0.15 \pm 0.32$           & 6, 15 \\
200826A                  & 0.748   & 0.65   & $41 \pm 2$            & $117 \pm 3$        & $0.63 \pm 0.03$           & 6, 15 \\
\enddata
\vspace{10pt}
{ 
\noindent \textbf{Notes:} The observed spectral parameters ($E_{p}$ and fluence) and $T_{90}$ are adopted from the referenced catalogs (e.g., Fermi/HEASARC, Swift Catalog) or literature. To ensure cosmological consistency, $E_{\gamma,\rm iso}$ for all sample bursts is uniformly derived by us in this work using standard cosmological parameters and appropriate $k$-corrections based on the adopted spectral models. For GRB 120422A, $E_p$ was estimated using the Yonetoku relation.

\vspace{5pt}
\noindent \textbf{References (Spectra \& Redshift):} 
(1) \citet{2006GCN..5264....1G}; 
(2) \citet{2022Natur.612..232Y};
(4) \citet{2013GCN.14771....1G};
(6) Fermi/HEASARC; 
(7) \citet{2020ApJ...893...46V};
(11) \citet{2021ApJ...916...89R}; 
(12) \citet{2024ApJ...969...26P};
(14) \citet{2017ApJ...850..161T}; 
(16) \citet{2005GCN..3479....1C};
(17) \citet{2008GCN..8369....1B}; 
(18) Swift Catalog; 
(19) \citet{2021ApJ...908...83T}; 
(20) \citet{2015GCN.18198....1G}; 
(21) \citet{2018GCN.23495....1S};
(22) \citet{2020AstL...46..783B}.

\vspace{5pt}
\noindent \textbf{References (SN \& KN classifications):} 
(3) \citet{2019MNRAS.486..672A}; 
(5) \citet{2013Natur.500..547T}; 
(8) \citet{2019MNRAS.490.4367T}; 
(9) \citet{2017ApJ...848L..12A};
(11) \citet{2021ApJ...916...89R}; 
(13) \citet{2024Natur.626..737L};
(15) \citet{2022ApJ...938...41D};
}
\end{deluxetable*}

\section{Methodology}
\label{sec:method}

\subsection{Feature Selection: Beyond the Duration Dichotomy}

Historically, the $T_{90}$ duration distribution has served as the primary criterion for classifying GRBs into long ($>2$\,s) and short ($<2$\,s) classes \citep{1993ApJ...413L.101K}. While this bimodality roughly traces the distinct physical progenitors---massive star collapsars and compact object mergers---recent observations of hybrid events have challenged this simple paradigm. Notable examples include the ``short'' collapsar GRB 200826A \citep{2021NatAs...5..911Z} and ``long'' mergers such as GRB 211211A \citep{2022Natur.612..223R} and GRB 230307A \citep{2024Natur.626..737L}. As discussed in \cite{2009ApJ...703.1696Z}, while the intrinsic duration of the central engine activity is physically distinct between the two populations, the observed $T_{90}$ is heavily contaminated by instrumental effects. Factors such as detector sensitivity, energy band, signal-to-noise ratio, and pulse shape can alter the observed duration by over an order of magnitude \citep{2013ApJ...765..116K, 2022ApJ...927..157M}, rendering $T_{90}$ alone an insufficient proxy for the progenitor system.

Multi-parameter classification schemes, such as the Type I (merger-origin) and Type II (collapsar-origin) framework proposed by \cite{2009ApJ...703.1696Z}, offer a more physically robust approach. However, this method relies on ancillary data---such as host galaxy morphology, offsets, and local burst environments---which are often difficult to obtain, particularly for faint or high-redshift hosts. Furthermore, even when available, these environmental properties can exhibit overlapping distributions, leading to potential ambiguities.

To overcome these limitations, we adopt a classification strategy based on three intrinsic prompt emission parameters: the rest-frame duration ($T_{90,z}$), the rest-frame peak energy ($E_{p,i} = E_p(1+z)$, where $E_p$ is the observed peak energy), and the isotropic equivalent energy ($E_{\rm iso}$). While these parameters require redshift measurements, they probe the central engine physics more directly, independent of the external environment. 

The selection of these input features is driven by distinct physical and methodological considerations:

\noindent \textbf{Physical Motivation:}
\begin{enumerate}
    \item \textbf{Intrinsic Timescale:} By using the rest-frame duration $T_{90,z} = T_{90}/(1+z)$, we correct for cosmological time dilation to directly probe the intrinsic activity scale of the central engine.
    
    \item \textbf{Radiative Diagnostics:} The $E_{p,i}-E_{\rm iso}$ correlation (Amati relation)  serves as a fundamental scaling law for collapsars \citep{2002A&A...390...81A}. Crucially, merger-origin GRBs are known outliers to this relation, typically exhibiting harder spectra for a given energy output. This separation in the spectral-energy plane provides a potent discriminator orthogonal to the time domain.
\end{enumerate}

\noindent \textbf{Methodological Advantage:}
\begin{enumerate}
    \setcounter{enumi}{2} 
    \item \textbf{Label Fidelity:} Unlike previous multi-parameter studies (e.g., \citealt{2010ApJ...725.1965L}) that relied on statistically classified samples, our training set is constructed strictly from ``gold-standard'' events with spectroscopically confirmed progenitors. This prevents the decision boundary from being contaminated by misclassified or hybrid events.
\end{enumerate}

\subsection{Machine Learning Algorithm: Support Vector Machine}

Given the high cost of obtaining ``gold-standard'' labels, our dataset is inherently limited in size ($N=24$). In this regime, complex models such as neural networks are prone to overfitting. We therefore employ the Support Vector Machine (SVM), a supervised learning algorithm specifically effective for small-sample classification problems \citep{Cortes1995}.

The SVM constructs a hyperplane  in the multi-dimensional parameter space that separates the two classes (collapsars and mergers) while maximizing the \textit{geometric margin}---the distance between the hyperplane and the nearest data points (support vectors). This margin maximization property ensures that the classifier is defined by critical boundary cases (e.g., short collapsars or long mergers) rather than the bulk distribution, yielding robust generalization to unseen data.

In this work, we implement a \textbf{linear kernel} SVM. While non-linear kernels (e.g., Radial Basis Function) can model more complex boundaries, we select the linear kernel for two primary reasons: 
(1) it minimizes the risk of overfitting, as our limited sample size ($N=24$) is insufficient to reliably constrain the boundary curvature required by non-linear kernels; and 
(2) it yields an explicit analytical formula (Equation \ref{eq:svm_index}) that enables direct physical interpretation of the feature weights, unlike the opaque decision rules of ``black box'' non-linear classifiers. The input features ($\log_{10} T_{90,z}$, $\log_{10} E_{p,i}$, $\log_{10} E_{\rm iso}$) are standardized (zero mean, unit variance) prior to training to ensure equal weighting during margin optimization. 

\begin{figure}[ht!]
    \centering
    \includegraphics[width=0.95\columnwidth]{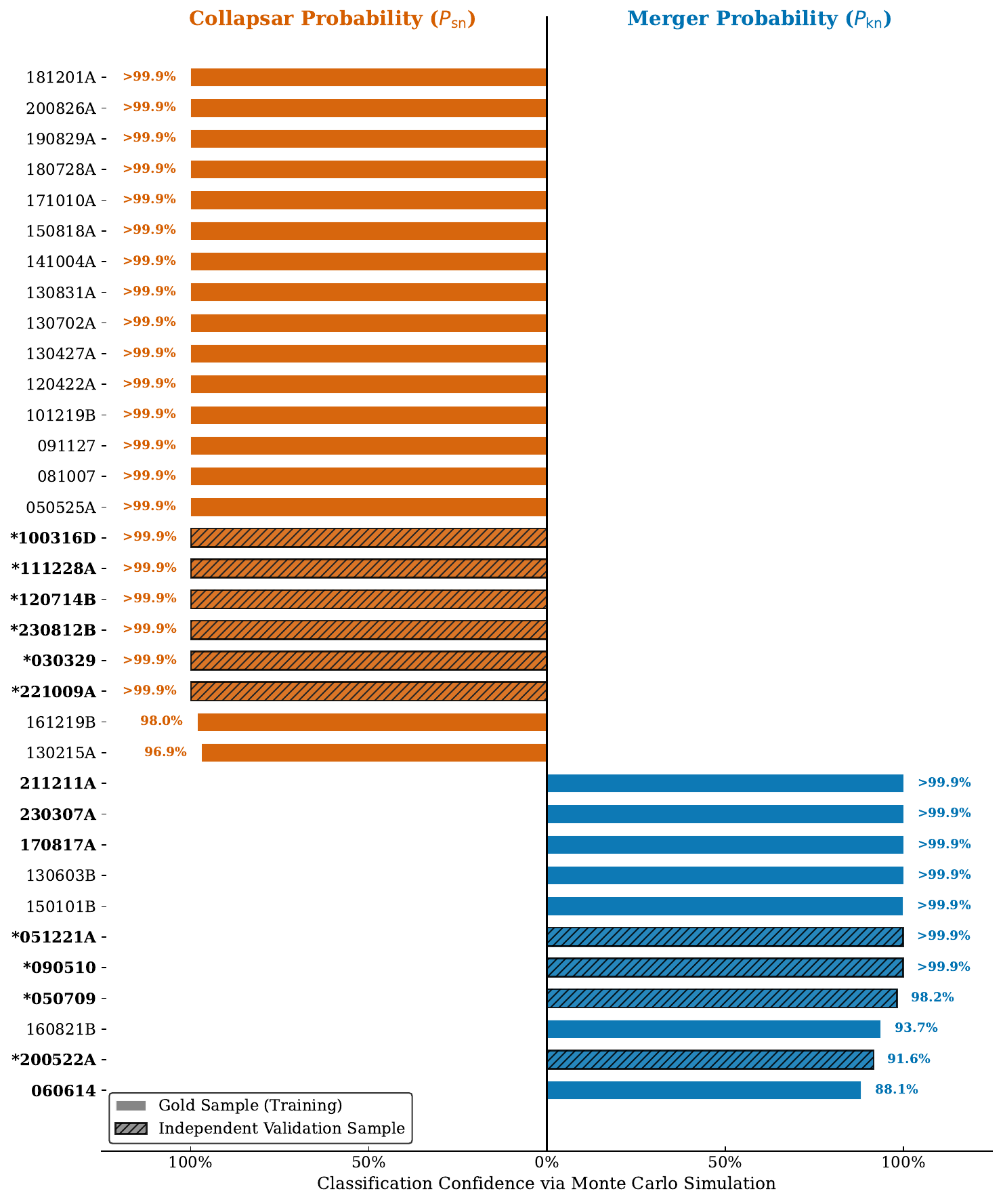}
    \caption{\textbf{Classification confidence and generalization.} A diverging bar chart showing the SVM classification probability for each burst. Bars extending to the right (blue) indicate high confidence for a merger origin, while bars to the left (red) indicate a collapsar origin. Hatched bars represent the 10 independent validation samples not included in the primary training set, demonstrating the model's robust generalization to historic events (e.g., GRB 050709) and extreme populations (e.g., GRB 221009A).}
    \label{fig:probability}
\end{figure}

\begin{figure}[ht!]
    \centering
    \includegraphics[width=0.95\columnwidth]{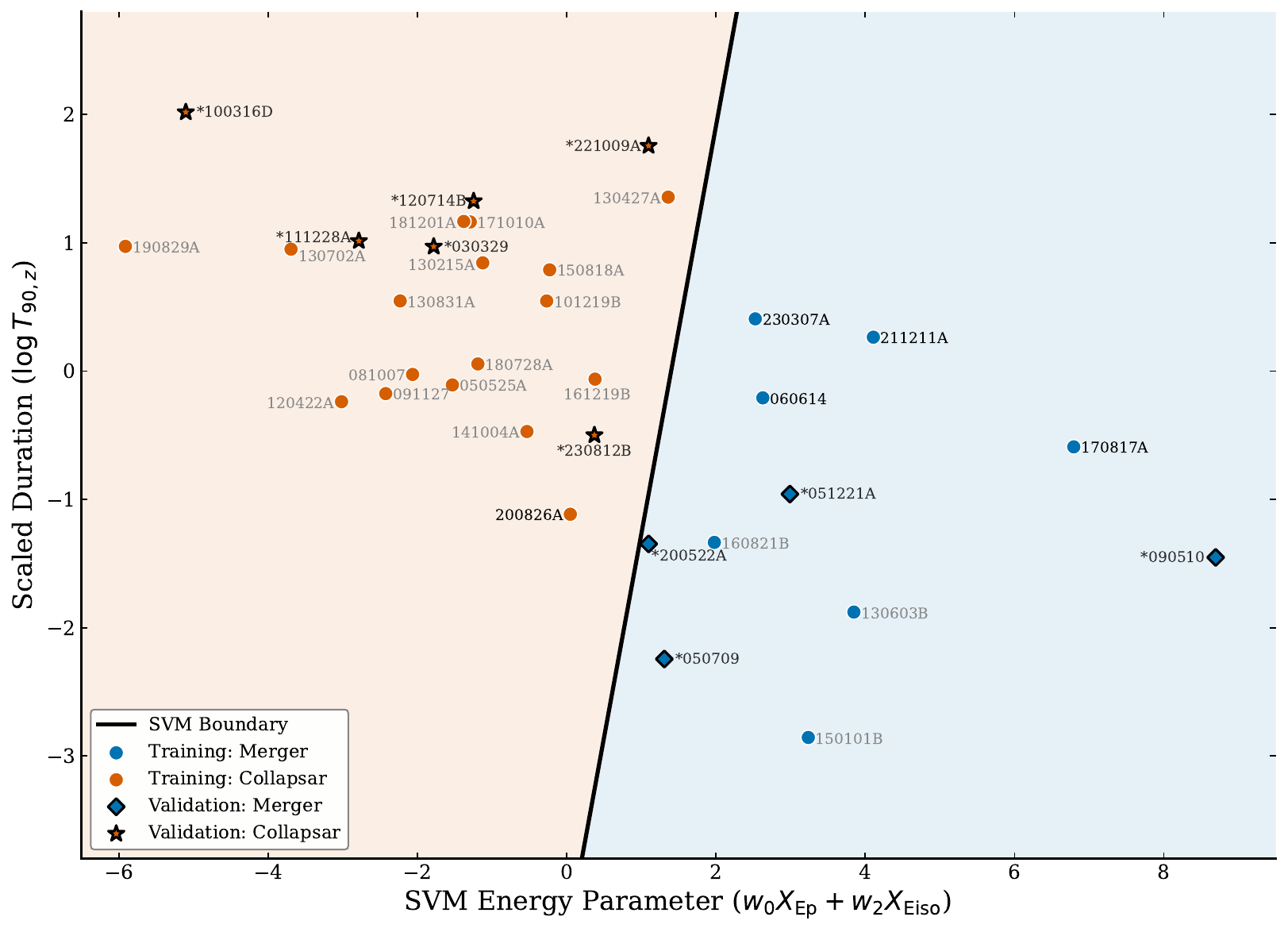}
    \caption{\textbf{The SVM classification plane.} Distribution of training and independent validation GRBs projected onto the 2D plane derived from the standardized $E_{p,i}$--$T_{90,z}$--$E_{\rm iso}$ parameter space. The solid black line indicates the optimal decision boundary established by the gold-standard training sample ($C=5$). Training events are represented by circles (blue for mergers, orange for collapsars). Independent validation events (Table \ref{tab:test_set}) are shown with enlarged markers and thick black edges: blue diamonds for mergers and orange stars for collapsars (asterisks denote validation GRBs). All validation cases are correctly classified into their respective physical regions, confirming the robustness and generalizability of the SVM model.}
    \label{fig:svm_boundary}
\end{figure}

\begin{figure}[ht!]
    \centering
    \includegraphics[width=0.95\columnwidth]{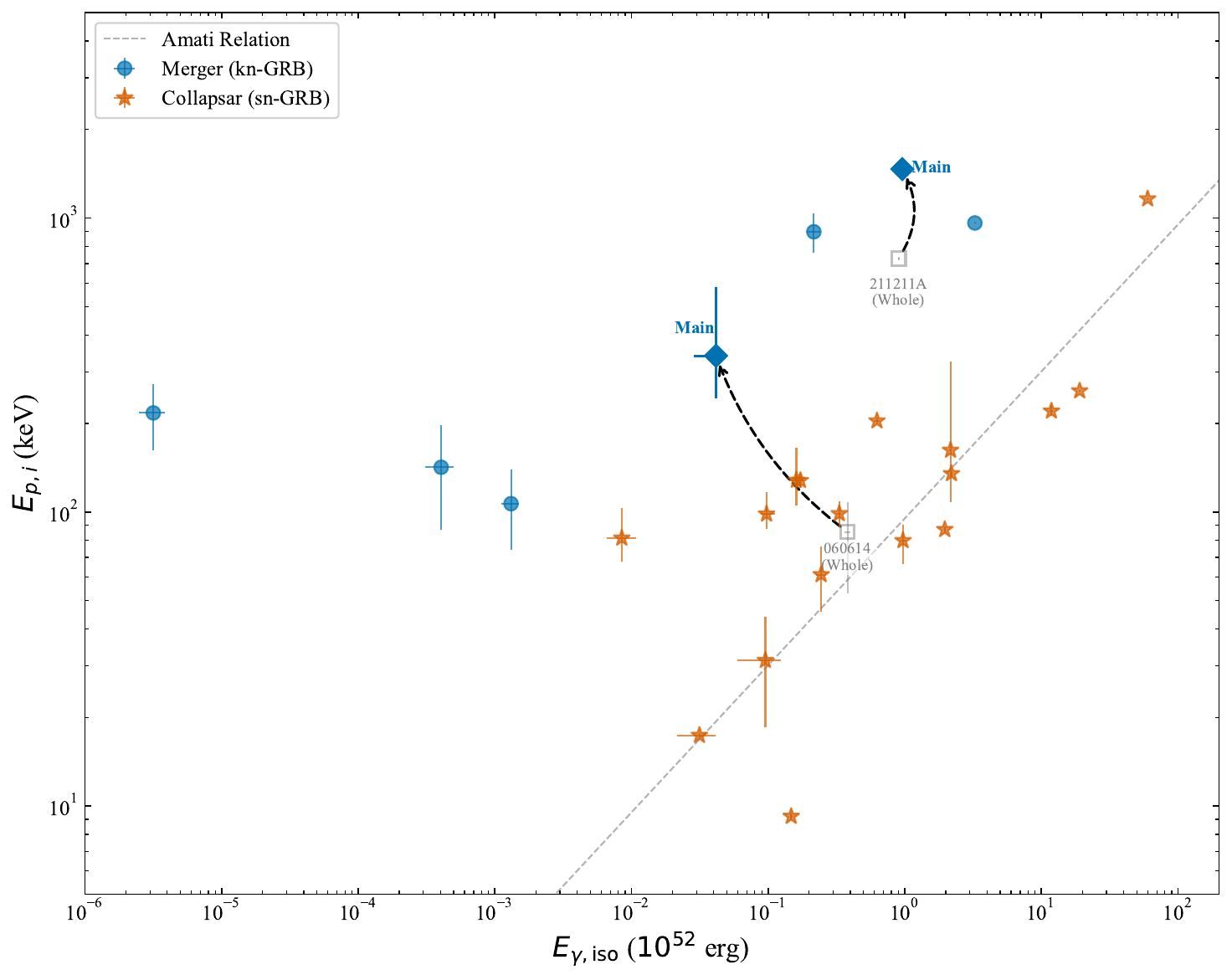}
    \caption{\textbf{Unveiling the progenitor physics.} The trajectory of hybrid bursts in the $E_{\rm iso}-E_{p,i}$ plane (Amati relation). Gray squares ``Whole'' represent the parameters derived from the time-integrated emission (including EE), which mimics collapsars. Blue diamonds (``Main'') represent the initial spike emission, showing a distinct migration back to the merger regime (blue circles) after removing the EE component. The dashed arrows illustrate this physical correction.}
    \label{fig:trajectory}
\end{figure}

\section{Results}
\label{sec:Results}

\subsection{Model Validation and Generalization}

To evaluate the robustness of our classifier given the limited sample size, we performed a Leave-One-Out Cross-Validation (LOOCV). In this validation scheme, the SVM model achieved an overall accuracy of \textbf{95.8\%} (23 out of 24 events correctly classified). The classification confidence for each event is presented in Figure \ref{fig:probability}. 

The only unstable case during cross-validation was GRB 200826A ($T_{90,z} \sim 0.65$\,s), which lies close to the geometric boundary. However, in the full model (Figure \ref{fig:svm_boundary}), this event is correctly recovered as a collapsar with a definitive score of $I_{\rm SVM} = -1.01$. This case is physically instructive: although its short duration mimics a merger by minimizing the penalty from the duration term, its spectral-energy properties fail to meet the high hardness threshold required to overcome the large negative intercept of the SVM boundary. The SVM correctly discerns that despite being short, it lacks the ``excess hardness'' characteristic of compact object mergers.

\subsection{The SVM Classification Boundary}

Using the full gold-standard training set, we derived the optimal decision hyperplane in the 3D parameter space. The resulting separation is visualized in Figure \ref{fig:svm_boundary}, which shows a clear separation between the merger (blue) and collapsar (red) populations. The resulting SVM decision boundary defines a classification index, $I_{\rm SVM}$, expressed as:
\begin{equation}
\label{eq:svm_index}
    I_{\rm SVM} \equiv 5.01 \log_{10} \left(\frac{E_{p,i}}{\rm keV}\right) - 1.25 \log_{10} \left(\frac{E_{\rm iso}}{10^{52}\,\rm erg}\right) - 0.34 \log_{10} \left(\frac{T_{90,z}}{\rm s}\right) - 12.90
\end{equation}
where events with $I_{\rm SVM} > 0$ are classified as mergers (kn-GRBs) and those with $I_{\rm SVM} < 0$ as collapsars (sn-GRBs). 

The coefficients of Equation \ref{eq:svm_index} reveal a clear hierarchy: the spectral peak energy ($E_{p,i}$) carries the highest discriminative weight. Analysis of the standardized feature weights reveals that the importance of $E_{p,i}$ is over 5 times greater than that of the duration ($T_{90,z}$). Furthermore, the isotropic energy ($E_{\rm iso}$) holds significant weight (only $\sim 1.5$ times lower than $E_{p,i}$).

This weighting hierarchy implies that the classification is fundamentally driven by the burst's location on the spectral-energy plane (i.e., the Amati relation) rather than its duration. This quantitative evidence confirms that intrinsic spectral hardness and energetics are the primary physical discriminators, whereas duration is a secondary feature.

This weighting scheme naturally resolves the ambiguity in the classification of hybrid bursts. A merger event can exhibit a long duration (e.g., GRB 211211A) yet be correctly identified ($I_{\rm SVM} > 0$) because its high intrinsic hardness ($E_{p,i}$) and moderate energy ($E_{\rm iso}$) outweigh the negative contribution from the duration term ($T_{90,z}$).

The robustness of this classification standard is validated by its successful generalization to events outside the training set. To rigorously test the model against observational extremes and avoid selection bias, we compiled an independent validation sample of 10 representative GRBs with secure population identities (Table \ref{tab:test_set}). The model unambiguously identifies the historic pre-\textit{Swift} burst GRB 030329 and the extraordinarily bright GRB 221009A as collapsars with secure confidence scores ($I_{\rm SVM} = -3.49$ and $-0.87$, respectively, both yielding $P_{\rm sn} > 0.999$). Furthermore, it successfully retrieves the HETE-2 detected short burst GRB 050709 ($I_{\rm SVM} = +0.61$) and the extreme high-energy GRB 090510 ($I_{\rm SVM} = +7.74$) as distinct mergers. Overall, the SVM classifier correctly categorizes all 10 validation events (6 collapsars and 4 mergers), yielding a validation accuracy of 100\% across this specific validation sample with no misclassifications. Notably, GRB 200522A, a strong kilonova candidate with marginal parameter values ($I_{\rm SVM} = +0.11$, $P_{\rm sn} = 0.084$), is still correctly identified as a merger, demonstrating the model's robustness even near the decision boundary. This confirms that the physical boundary captured by the SVM is intrinsic and robust against specific instrument capabilities or extreme parameter ranges.  We propose $I_{\rm SVM}$ as a practical tool for identifying merger candidates in the multi-messenger era, particularly when gravitational wave data are absent.

\begin{deluxetable*}{lccccccccc}
\tablecaption{Independent Validation Sample of 10 GRBs with Secure Progenitor Identities \label{tab:test_set}}
\tablewidth{0pt}
\tabletypesize{\small}
\tablehead{
    \colhead{GRB} &
    \colhead{$z$} &
    \colhead{$T_{90,z}$ (s)} &
    \colhead{$E_{p,i}$ (keV)} &
    \colhead{$E_{\rm iso}$ ($10^{52}$ erg)} &
    \colhead{$I_{\rm SVM}$} &
    \colhead{$P_{\rm sn}$} &
    \colhead{Pred.} &
    \colhead{Refs.} &
    \colhead{Note}
}
\startdata
\cutinhead{Collapsars (SN-GRBs)}
221009A & 0.151  & $283.1  \pm 5.9$   & $2170.8 \pm 84.0$    & $1200.00 \pm 100.00$ & $-0.87$ & $>0.999$ & SN & 1--3    & (i)   \\
030329  & 0.1685 & $53.8   \pm 3.3$   & $120.4  \pm 2.3$     & $2.20    \pm 0.03$   & $-3.49$ & $>0.999$ & SN & 4--6    & (ii)  \\
230812B & 0.360  & $2.40   \pm 0.07$  & $435.2  \pm 2.7$     & $7.21    \pm 0.05$   & $-0.88$ & $>0.999$ & SN & 7, 8    & (iii) \\
120714B & 0.3984 & $113.7  \pm 24.3$  & $60.0   \pm 25.9$    & $0.050   \pm 0.008$  & $-3.06$ & $>0.999$ & SN & 9, 10   & (iv)  \\
111228A & 0.7163 & $59.0   \pm 3.1$   & $73.8   \pm 30.9$    & $1.96    \pm 0.20$   & $-4.51$ & $>0.999$ & SN & 5, 10, 11 & (v)   \\
100316D & 0.0591 & $492.8  \pm 415.1$ & $10.2   \pm 9.2$     & $0.050   \pm 0.009$  & $-7.14$ & $>0.999$ & SN & 5, 12   & (vi)  \\
\cutinhead{Mergers (KN/Short-GRBs)}
090510  & 0.903  & $0.32   \pm 0.02$  & $17117.5 \pm 1263.8$ & $3.91    \pm 0.09$   & $+7.74$ & $<0.001$ & KN & 13, 15  & (vii)  \\
051221A & 0.546  & $0.91   \pm 0.13$  & $621.5  \pm 143.8$   & $0.24    \pm 0.02$   & $+1.88$ & $<0.001$ & KN & 14, 15  & (viii) \\
050709  & 0.1606 & $0.060  \pm 0.009$ & $96.3   \pm 20.9$    & $0.003   \pm 0.0003$ & $+0.61$ & $0.018$  & KN & 4, 16   & (ix)   \\
200522A & 0.554  & $0.40   \pm 0.05$  & $120.8  \pm 12.1$    & $0.011   \pm 0.001$  & $+0.11$ & $0.084$  & KN & 10      & (x)    \\
\enddata
\vspace{8pt}
{\footnotesize 
\noindent \textbf{Notes:} None of the validation events are included in the primary training set (Table 1). $T_{90,z}$ and $E_{p,i}$ are rest-frame values derived via $(1+z)^{-1}$ and $(1+z)$ corrections, respectively. $I_{\rm SVM}$ is the deterministic geometrical score from central parameter values. $P_{\rm sn}$ is the collapsar probability via $10^4$ Monte Carlo simulations over the observational error space.\\[1ex]
\textbf{Selection Reasons:} 
(i)~Extreme $E_{\rm iso}$ outlier; secure SN association. 
(ii)~Archetypal bright local GRB-SN; historical benchmark. 
(iii)~Recent high-quality GRB-SN; expands temporal coverage. 
(iv)~Typical cosmological SN-GRB; standard parameter range. 
(v)~Standard long GRB with secure SN; baseline validation. 
(vi)~Extreme low-luminosity end; tests parameter boundary. 
(vii)~Extreme $E_{p,i}$ outlier; tests high-energy boundary. 
(viii)~Standard bright short GRB; baseline merger validation. 
(ix)~First short GRB with optical counterpart; historical benchmark. 
(x)~Marginally classified merger; tests decision boundary robustness.\\[1ex]
\textbf{References (Spectra \& Redshift):} 
(1)~\citet{2023ApJ...949L...7F}; 
(2)~\citet{2023ApJ...946L..26A}; 
(4)~\citet{2009ApJ...703.1696Z}; 
(5)~\citet{2022ApJ...938...41D}; 
(7)~\url{https://heasarc.gsfc.nasa.gov}; 
(9)~\citet{2015ApJS..218...13Y}; 
(11)~\citet{2023ApJS..266...31L}; 
(12)~\citet{2018ApJ...866...97B}; 
(13)~\citet{2025ApJ...991..209Z}; 
(14)~\citet{2006ApJ...650..261S}.\\[1ex]
\textbf{References (SN \& KN Classifications):} 
(3)~\citet{2023ApJ...946L..22F}; 
(6)~\citet{2003ApJ...591L..17S}; 
(8)~\citet{2024ApJ...960L..18S}; 
(10)~\citet{2023MNRAS.524.1096L}; 
(15)~\citet{2014ARAA..52...43B}; 
(16)~\citet{2016NatCo...712898J}.
}
\end{deluxetable*}

\section{Discussion}
\label{sec:discussion}

\subsection{The Primacy of Spectral Hardness}
The most striking feature of our SVM classification index ($I_{\rm SVM}$) is the dominant weight assigned to the rest-frame peak energy ($E_{p,i}$).
Physically, we can visualize the classification boundary in the $E_{p,i}-E_{\rm iso}$ plane by examining the projection where the secondary term $T_{90,z}$ is held constant at its sample median. From Equation \ref{eq:svm_index}, setting $I_{\rm SVM}=0$ yields a boundary slope of:
\begin{equation}
    \frac{d(\log E_{p,i})}{d(\log E_{\rm iso})} = \frac{1.25}{5.01} \approx 0.25
\end{equation}
It is intriguing to note that this boundary slope ($\propto E_{\rm iso}^{0.25}$) is significantly \textit{shallower} than the classic Amati relation ($\propto E_{\rm iso}^{0.5}$) historically established for long GRBs. 

This discrepancy suggests a quantitative ``hardness ceiling'' for the collapsar population, strictly defined by the $E_{\rm iso}^{0.25}$ dependence: while the bulk of long GRBs (predominantly collapsars) clusters around the steeper 0.5 power law, the critical boundary separating them from mergers rises much more slowly. This effectively restricts pure collapsars to a narrower spectral range at high energies. In contrast, mergers inhabit the regime above this boundary, characterized by an ``excess'' of hardness relative to their energy budget. This supports the hypothesis that merger-driven jets are intrinsically harder and that this spectral signature is their most defining trait, far more reliable than the duration of the prompt emission.

\subsection{Decoupling the Main Engine from Extended Emission}
Hybrid events—such as the kilonova-associated long GRBs 060614 and 211211A—have historically obscured the phenomenological boundary between the two populations. Our analysis reveals that the confusion largely stems from the inclusion of the soft Extended Emission (EE) in the global parameter estimation. As shown in Figure \ref{fig:trajectory}, when the EE component is removed and parameters are derived solely from the initial main spike, these hybrid bursts migrate from the collapsar region back into the merger domain defined by our SVM boundary.

This suggests that the main spike and the EE may originate from distinct physical processes. The main spike tracks the primary relativistic jet launched during the merger, carrying the imprint of the central engine's hardness. The EE, conversely, may arise from secondary mechanisms such as fallback accretion or magnetar spin-down \citep{2008MNRAS.385.1455M, 2022Natur.612..223R}, which mimic the timescale of collapsars but lack their spectral-energy correlation properties. Consequently, we argue that a reliable physical classification must isolate the prompt main spike, treating the EE as a distinct phenomenology that contaminates the progenitor signal.

\subsection{Applicability and Selection Effects}
While our ``gold-standard'' merger sample is currently limited to $z \lesssim 0.4$, the discovered classification boundary appears to be structural rather than an artifact of redshift bias. Since our SVM is trained on \textit{rest-frame} parameters ($E_{p,i}$, $E_{\rm iso}$, $T_{90,z}$), the cosmological $k$-correction and time dilation are mathematically accounted for, minimizing the direct impact of redshift evolution. 
Our methodology aligns with the $\epsilon$-parameter proposed by \citet{2010ApJ...725.1965L}, which first identified the distinct tracks of mergers and collapsars in the $E_{p,i}-E_{\rm iso}$ plane. However, by integrating duration ($T_{90,z}$) and utilizing a supervised learning algorithm to optimize the weights, our $I_{\rm SVM}$ index provides a more rigorous and quantitative separation boundary.

It is important to note that our classifier is strictly calibrated for classical GRBs. Low-luminosity GRBs (ll GRBs) associated with SNe (e.g., GRB 060218) are excluded from our scope. These events, likely driven by shock breakouts rather than highly relativistic jets \citep{2006Natur.442.1008C}, follow a different physical track and should be treated as a separate population.

Finally, the classification tool presented here paves the way for high-precision GRB cosmology. As highlighted in the Introduction, sample impurity has been a major source of systematic error in previous cosmological studies. By applying the $I_{\rm SVM}$ criterion to the vast archival data from \textit{Swift}, \textit{Fermi}, and future observations from \textit{SVOM}, we can effectively excise ``imposter'' merger events from the long-GRB population. Our future work will focus on constructing a ``cosmology-grade'' sample of pure collapsars to recalibrate the luminosity correlations (e.g., the Amati relation).If merger contamination contributes to the observed intrinsic scatter, purification using $I_{\rm SVM}$ may significantly reduce the dispersion in the Hubble diagram. This would enable GRBs to serve as more reliable standardizable candles for probing the expansion history up to $z \sim 15$.

\begin{acknowledgments}
We thank Z.-Y. Peng for helpful discussions. This work was supported by the National Natural Science Foundation of China (grant Nos. 12494575 and 12273009), the Shandong Provincial Natural Science Foundation (grant Nos. ZR2025MS16, ZR2025MS47 and ZR2025QC25), the China Postdoctoral Science Foundation (grant No. 2025M783226), and the Key R\&D Program of Jining City, Shandong Province, China (grant No. 2025ZDZP033). F. Lyu is supported by the National Natural Science Foundation of China (grant No. 12403042).
\end{acknowledgments}

%

\vspace{5mm}
\facilities{Swift(BAT), Fermi(GBM), Konus-Wind, HETE-2, LIGO, Virgo}





\bibliography{sample631}{}
\bibliographystyle{aasjournal}



\end{document}